\documentclass[12pt]{iopart}

\usepackage{graphicx} 
\usepackage{dcolumn}  
\usepackage{bm}       
\newcommand{\be}{\begin{equation}}
\newcommand{\ee}{\end{equation}}
\newcommand{\ba}{\begin{eqnarray}}
\newcommand{\ea}{\end{eqnarray}}
\newcommand{\bse}{\begin{subequations}}
\newcommand{\ese}{\end{subequations}}
\newcommand{\M}{{\cal {M}}}

\newcommand{\CZ}{{\cal {Z}}}

\newcommand{\Mvir}{M_{\textrm{\tiny{vir}}}}

\newcommand{\dd}{\textrm{d}}

\begin{document}

\title[Stellar polytropes and NFW halo models]{Stellar polytropes and Navarro-Frenk-White halo models: comparison with observations}

\author{Jes\'us Zavala$^1$\footnote{Visiting Researcher, Max Planck Institute for
Astrophysics, Munich, Germany}, Dar\'\i o N\'u\~nez$^1$, Roberto
A. Sussman$^1$, Luis G. Cabral-Rosetti$^2$ and Tonatiuh
Matos$^3$\footnote{Visiting Researcher, CIAR Cosmology and Gravity
Program, Department of Physics and Astronomy, University of British
Columbia, Vancouver, British Columbia, Canada, V6T 1Z1}}

\address{$^1$ Instituto de Ciencias Nucleares Universidad Nacional Aut\'onoma de M\'exico A. P. 70-543,  M\'exico 04510 D.F., M\'exico}

\eads{\mailto{nunez@nucleares.unam.mx}, \mailto{sussman@nucleares.unam.mx}, \mailto{jzavala@nucleares.unam.mx}}

\address{$^2$ Departamento de Posgrado, Centro Interdisciplinario de Investigaci\'on y Docencia en Educaci\'on T\'ecnica (CIIDET), Av. Universidad 282 Pte., Col. Centro, A. Postal 752, C. P. 76000, Santiago de Quer\'etaro, Qro., M\'exico.}

\ead{lgcabral@ciidet.edu.mx}

\address{$^3$ Departamento de F\'\i sica, Centro de Investigaci\'on y de Estudios Avanzados del IPN A. P. 14-740,  M\'exico 07000 D.F., M\'exico}

\ead{tmatos@fis.cinvestav.mx}

\begin{abstract}

Motivated by the possible conflict between the
Navarro-Frenk-White(NFW) model predictions for the dark matter
contents of galactic systems and its correlation with baryonic surface
density, we will explore an alternative paradigm for the description
of dark matter halos. Such an alternative emerges from Tsallis'
non-extensive thermodynamics applied to self-gravitating systems and leads
to the so-called ``stellar polytrope'' (SP) model. We consider that
this could be a better approach to real structures 
rather than the isothermal model, given the fact that the first one takes into 
account the non-extensivity of energy and entropy present in these type of 
systems characterized by long-range interactions. We compare a halo based on 
the Navarro-Frenk-White (NFW) and one which follows the SP description. 
Analyzing the dark matter contents estimated by means of global
physical parameters of 
galactic disks, obtained from a sample of actual galaxies, with the ones of the 
unobserved dark matter halos, we conclude that the SP model is favored over 
the NFW model in such a comparison. 

\end{abstract}

\noindent{\it keywords\/}: dark matter, galaxy dynamics


\section{Introduction}

The standard statistical mechanical treatment for self--gravitational
systems is provided by the micro-canonical ensemble, leading to a regime dominated by 
gravity and characterized by well known effects such as the non--extensive nature of
energy and entropy, negative heat capacities and the so--called gravothermal
instability. An alternative formalism that allows also non-extensive
forms for entropy and energy under simplified assumptions, has been developed by Tsallis 
\cite{Tsallis} and
applied to self--gravitating systems \cite{PL,TS1,TS2} under the assumption of a
kinetic theory treatment and a mean field approximation. As opposed to the
Maxwell--Boltzmann distribution that follows as the equilibrium state associated with
the  usual Boltzmann--Gibbs entropy functional, the Tsallis' functional
yields  as equilibrium state the ``stellar polytrope'',
characterized by a polytropic equation of state with index $n$ (see
equation (\ref{eq:edo})). The stellar polytrope (SP)
yields a Maxwell--Boltzmann distribution function (the isothermal sphere) in the
limit $n\to\infty$. This index is related to the
``non--extensivity'' parameter $q$ of Tsallis entropy functional,
so that the ``extensivity'' limit $q\to 1$ corresponds to the
isothermal sphere.

Although the self--gravitating  collisionless and virialized gas
that makes up galactic halos is far from the state of
gravothermal catastrophe, it is reasonable to assume that it is
near some form of relaxation equilibrium characterized by
non--extensive forms of entropy and energy. Admitting that the SP model follows
from an idealized approach based on kinetic theory and an isotropic 
distribution function, it is interesting to verify empirically if the
structural parameters of the halo gas can be adjusted to those of SPs, the equilibrium 
states under Tsallis' formalism. 

On the other hand, high precision N--body  numerical simulations based
on Cold Dark Matter (CDM) models, 
perhaps the most powerful method available for understanding gravitational
clustering, lead to the famous results of Navarro,
Frenk and White (NFW model) \cite{NFW} that predicts density and velocity profiles which are 
roughly consistent with observations; however, it also predicts a cuspy 
behavior at the center of galaxies that is not observed in most of the rotation
curves of dwarf and LSB galaxies \cite{Blok1,Blok2,Bin,B-S,B-O,Bosma3}. The 
significance of this discrepancy with observations is still under dispute, 
leading to various theoretical alternatives, either within the thermal
paradigm (self--interacting \cite{self} and/or ``warm''
\cite{warm} dark matter, made of lighter particles), or non--thermal
dark matter models (with real \cite{DMe1,DMe2,DMe3,DMe4,DMe5} or complex \cite{Ruffini}
scalar fields, axions, etc), and none of these alternatives
is free of controversy. However, for larger scales
like galaxy clusters, the NFW paradigm is strongly supported by
observations. A recent analysis \cite{Pointe} of ten nearby clusters
using the X-ray satellite XMM-Newton has shown that clusters have
indeed a cusped nature (see also \cite{Voigt} for a similar study
using the Chandra satellite, and also \cite{Zapa} for a particular
case study). Therefore, the CDM model of
collision--less WIMPs remains as a viable model to account for dark matter
in galactic halos, provided there is a mechanism to explain the
discrepancies of this model with observations in the center of
galaxies.

Since gravity is a long--range interaction and virialized self--gravitating 
systems are characterized by non--extensive forms of entropy and energy, it is
reasonable to expect that the final configurations of halo
structure predicted by N--body simulations must be, somehow,
related with states of relaxation associated with non--extensive
formulations of Statistical Mechanics; therefore, a comparison between the SP
and NFW models is both, possible and interesting.
 
A theoretical comparison between the NFW model an the SP model is
established in \cite{Tsallis1}. The aim of that paper was to verify which parameters of the stellar polytropes 
provide a suitable description of the halo that resembles the one
that emerges from the NFW model. A convenient criterion was established 
for a best fit of the SP model to the NFW profiles followed by finding
the stellar polytrope whose central density, $\rho_c$, central velocity
dispersion, $\sigma_c$, and polytropic index, $n$, yields the same virial
mass, total energy and maximal velocity of a given NFW halo model.
Considering halo virial masses in the range $10^{12}-10^{15}\,M_\odot$, it was found that 
the best fit to NFW profiles at all scales is given by polytropic indices close to
$n\approx 4.8$, leading to an empirical estimation of Tsallis non--extensive
parameter: $q\approx 1.3$. We will describe with more detail the main idea and results of that
work in the present paper.

The main purpose of our analysis is to make a dynamical analysis of
two halo models, one based on the NFW paradigm, and other based on
the SPs derived from Tsallis' non--extensive
thermodynamics, compare them and test both with observational
results coming from a sample of disk galaxies.

The paper is organized as follows: in section II we provide the equilibrium
equations of SPs and briefly summarize their connection
to Tsallis' non--extensive entropy formalism. In section III we discuss the
dynamical variables of NFW halos, while in section IV 
we describe a procedure to compare a polytropic halo with an NFW one and obtain numerically
the parameters that characterize such polytropic halos. Section V
deals theoretically with the dynamical consequences 
due to the formation of the galactic disk within the halo. In section VI we describe
the comparison with observations as well as the method
we used to make such a comparison. A summary of our results is given in
section VII.

\section{Tsallis' entropy and stellar polytropes}

For a phase space given by $({\bf{r}},\,{\bf{p}})$, the kinetic
theory entropy functional associated with Tsallis' formalism
is \cite{PL,TS1,TS2}:
\begin{equation}
S_{q}\ =\ -\frac{1}{q-1}\,\int
{(f^{q}-f)\,{d}^{3}{\bf{r}}\,%
{d}^{3}{\bf{p}}},  \label{q_entropy}
\end{equation}
where $f$ is the distribution function and $q>1$ is a real number. In the
limit $q\rightarrow 1$, the functional (\ref{q_entropy}) leads to the usual
Boltzmann--Gibbs functional, corresponding to the isothermal sphere. The condition 
$\delta \,S_{q}=0$ leads to the distribution function that corresponds to the 
SP model characterized by the equation of state:
\begin{equation}
p\ =\ K_n\,\rho ^{1+1/n}, \label{eq:edo}
\end{equation}
where $K_n$ is a function of the polytropic index $n$, and can be expressed
in terms of the central parameters:
\begin{equation} K_n=\frac{{\sigma_c}^2}{{\rho_c}^{1/n}}
\label{eq:Kn}\end{equation}
The polytropic index, $n$, is related to the Tsallis' parameter $q>1$ by:
\begin{equation}\label{n}
n\ =\ \frac{3}{2}+\frac{1}{q-1}
\end{equation}

The standard approach for studying spherically symmetric hydrostatic
equilibrium in stellar polytropes follows from inserting (\ref{eq:edo}) into
Poisson's equation, leading to the well known Lane--Emden equation~\cite{B-T}:
\begin{eqnarray}
\frac{1}{x^2}\,\frac{{d}}{{d}
x}\left(x^2\,\frac{{d}\,\theta}{%
{d} x}\right)+\theta^n \ = \ 0,  \label{eq:LE}
\end{eqnarray}
with
\begin{eqnarray} 
\theta \ &=& \ \left(\frac{\rho}{\rho_c}\right)^{1/n},
\\ x \ &=& \ \frac{r}{r_0},\quad r_0^{-2} \ = \
\frac{4\pi G\,
\rho_c}{\sigma_c^2
}, \quad \sigma_c^2 \ = \ \frac{p_c}{\rho_c},\\
&G \ &= \ 4.297\times 10^{-6}\,\frac{\left(\rm{km/sec}\right)^2}
{\rm{M}_\odot/\rm{kpc}},
\end{eqnarray}
where, as mentioned above, $\sigma_c$ and $\rho_c$ are the central velocity dispersion
and central mass density respectively; $G$ is the gravitational 
constant in appropriate units. Notice that the velocity dispersion is a
measure of the kinetic temperature of the gas by means of the relation:
$\sigma_c^2=k_{_B}\,T_c/m$, with $k_{_B}$ the Boltzmann's
constant, and that we are using a normalization for $r_0$, which
differs from the usual one by a factor $1/(n+1)$. We find it more
convenient to consider instead of equation (\ref{eq:LE}) the following set
of equivalent equilibrium equations:
\begin{eqnarray}
\frac{\dd \M}{\dd x} && = \ x^2\,\CZ,
\nonumber \\
\frac{\dd\CZ}{\dd x} && = \
-\frac{n}{n+1}\,\frac{\M\,\CZ^{1-1/n}}{x^2}, \label{eq_ZMx}
\end{eqnarray}
where $\M$ and $\CZ$ relate to $M,\,\rho$
(mass and mass density at radius $r$) by:
\begin{eqnarray}
\M && = \ \frac{M}{4\pi \rho_c\,r_0^3}, \qquad \CZ
\ = \ \frac{\rho}{\rho_c}
  \label{pars}
\end{eqnarray}
Notice that in the limit $n\to\infty $, equations
(\ref{eq_ZMx}) become the equilibrium equations of the isothermal sphere.

Once the system (\ref{eq_ZMx}) has been integrated numerically,
the velocity profile derived from the virial theorem takes the form:
\begin{equation}V^2(x) \ = \ \sigma_c^2\,\frac{\M}{x}
\label{rotvel}\end{equation}
The radial distance $r$ in kpc and enclosed mass
$M(r)$ in solar masses are given (from equation (\ref{pars})) in terms of $x$ and $\M$
by:
\begin{eqnarray}
&r/{\rm{kpc}} & = \ 0.004220\,\, \frac{\sigma_c}{\rm{km/sec}}\,\,
\left(\frac{\rm{M}_\odot/\rm{pc}^3}{\rho_c}\right)^{1/2}\,\,x,
\nonumber
\\
&& \\
&M/{\rm{M}_\odot} & = \ 944.97\,\,
\left(\frac{\sigma_c}{\rm{km/sec}}\right)^3\,\,
\left(\frac{\rm{M}_\odot/\rm{pc^3}}{\rho_c}%
\right)^{1/2}\,\,\M \nonumber
\label{physvars}\end{eqnarray}
Another important dynamical quantity is the total energy of the
stellar polytrope~\cite{TS1}:
\ba
E \ &=& \ K+U\nonumber
\\&=& \ \frac{3}{2}{\int_0}^{r}\,4\pi r^2 P(r)\,dr -{\int_0}^{r}{dr
\frac{GM(r)}{r}\frac{dM(r)}{dr}},\nonumber\\  \label{eq:E}
\ea
leading to
\begin{equation} E_{poly} = -\, \frac{1}{n - 5}\Big[
\frac{3}{2}\,\frac{G\,M^2}{r}
-\Big(\frac{\rho}{\rho_c}\Big)^{1/n}\,\sigma_c\Big(\frac{3}{2}
(n + 1) M_v-(n - 2)4\,\pi\,{r}^3\,\rho_v \Big)\Big],  \label{eq:E_poly}
 \end{equation}
which must be evaluated at a fixed, but arbitrary, value of $r$ marking a
cut--off scale.

Despite the valid general approach presented here it should be
emphasized that there is confusion on the literature regarding the
connection between $n$ and $q$ (equation (\ref{n})). The difference between different authors in this matter is mainly
connected to the ambiguity on a proper definition of the one-particle
distribution $f({\bf{r}},{\bf{p}})$ on terms of the probability
$p({\bf{r}},{\bf{p}})$. In the ``old'' Tsallis' formalism, the
relation is $f\propto p$ which for self-gravitating systems reveals
that the extremum state for the entropy gives: 
$f\propto (\Phi_0-v^2/2+\Phi({\bf{r}}))^{1/(q-1)}$ where
$\Phi({\bf{r}})$ is the gravitational potential (see for example
\cite{TS1}). Such a distribution then leads to equation (\ref{n}). The
``new'' Tsallis' formalism introduced in \cite{TsaPla} establishes the
``choice'': $f\propto p^q$; under this definition the same treatment as
in \cite{TS1} leads to \cite{Taru_new}: $f\propto
(\Phi_0-v^2/2+\Phi({\bf{r}}))^{q/(q-1)}$ and then to $n=1/2+1/(1-q)$
(see also \cite{Taru_05}). Another possibility exists for the
connection between $n$ and $q$ using a different approach for the
treatment of self-gravitating systems \cite{Lima} which leads to 
$n=(5-3q)/2(1-q)$. There is still not a final word on which of these
three possibilities is the most correct.

The first two are closely related by a so called {\it duality
transformation} $q\leftrightarrow 1/q$ in the distribution function
which seems to indicate that the thermodynamics properties in the new
formalism ($f\propto p^q$) can be translated into the old formalism
($f\propto p$) at least for the microcanonical ensemble description
(see \cite{Taru_new}). 

The differences between the third description and the first two are more
complicated and out of the scope of this article; in
\cite{Hansen_Tsallis}, the authors present an analytical and numerical
study for the velocity distribution functions of self-gravitating
systems and they claim that it is in fact this third possibility the
one that better agrees with their results.

Since there is yet no unique general agreement on which connection
between $n$ and $q$, and the corresponding description leading to it,
is the correct one, we will use the one presented in equation
(\ref{n}), considering that for the purposes of this work, the
different expressions between $n$ and $q$ are just different ways of
parameterizing the SPs and therefore, using one or the other will not lead to qualitative
changes in the subsequent results. However it is important
to be clear that this might not be true and a deeper analysis is
needed to be sure about this issue; at least, as is clear for the different
formulas connecting $n$ and $q$, the value for $q$ found and reported
in Table 1 is absolutely dependent on our particular choice for the
description leading to equation (\ref{n}).

\section{NFW halos}

It is important to say that the NFW density profile \cite{NFW}
that we will describe in the present section and that will be used in
this paper, is nowadays believed not to be the ultimate answer to the
actual density profile of dark matter halos. 

As was mentioned in the Introduction, the NFW density profile shows a
cuspy behavior in the center,
in fact, for this region, it has a power law dependence that goes as
$\rho\sim r^{-1}$ whereas $\rho\sim r^{-3}$ for the outer regions; such
features can easily be seen in equation (\ref{rho_nfw}). Recent
numerical simulations have shown that different profiles can adjust
better to the simulations (see for example \cite{NT} that propose a
model where $\rho\sim r^{-0.75}$ in the center, see also \cite{Sto} for
a recent numerical analysis and the profile model discussed
there). Also, semianalytical models \cite{Austin, Dehnen, Hansen}
using a power law dependence for the phase space density:
$\rho/\sigma^3\propto r^{-\alpha}$ (which holds for simulated dark
matter halos), where $\sigma$ is the velocity dispersion, together
with the Jeans equation have been established a model for a better
understanding of the dynamics of halos; predicting in particular power
law dependences of the density profile going as $\rho\sim r^{-0.8}$ in
the central part and $\rho\sim r^{-3.44}$ in the external region. 

However, to simplify our analysis we will use the NFW profile as a
test model due to its well known analytical expressions, presented in
the following paragraphs, leaving the analysis using more accurate
models as a possible future work.

NFW numerical simulations yield the following expression for the density
profile of virialized  galactic halo structures \cite{NFW,Mo}:
\begin{equation}\label{rho_nfw}
\rho_{_{\mathrm{NFW}}}=\frac{\delta_0\,\rho_0}{y\,\left(1+y
\right)^2},   \label{rho_NFW}
\end{equation}
where:
\begin{eqnarray} \delta_0 &&=
\frac{\Delta\,c_0^3}{3\left[\ln\,(1+c_0)-c_0/
(1+c_0)\right]},\label{delta0}\\\nonumber\\
\rho_0&&=\rho_{\mathrm{crit}}\,\Omega_0\,h^2 = 253.8 \, h^2\,\Omega_0\,
\frac{M_\odot}{\rm{kpc}^3},\label{rho0}\\
y &&= c_0\frac{r}{r_v},\label{y}
\end{eqnarray}
where $\Omega_0$ is the ratio of the total density to the critical 
density of the Universe, being $1$ for a flat Universe.
Notice that we are using a scale parameter, $y$, that is different from that
of the stellar polytropes, $x$. The NFW virial radius, $r_v$, is given in terms
of the virial mass, $M_v$, by  the condition that average halo density
is proportional to the  cosmological density $\rho_0$:
\begin{equation}\Delta\, \rho_0 = \frac{3\,M_v}{4\,\pi\,r_v^3},
\label{rvir}\end{equation}
where $\Delta$ is a model--dependent numerical factor (for a
$\Lambda$CDM model we have $\Delta\sim 100$ at $z=0$ \cite{LoHo}).
We remark that this last relation between the mass and
virial radius in terms of cosmological parameters, equation (\ref{rvir}), is valid
for any halo model. The concentration parameter $c_0$ can be expressed in terms of
$M_v$ by \cite{JZ,c0}:
\begin{equation} c_0 \approx 62.1 \times
\left(\frac{M_v\,h}{M_\odot}\right)^{-0.06} \label{eq:c0}
\end{equation}
hence all quantities depend on a single free parameter $M_v$. The mass
function and circular velocity follow from (\ref{rho_NFW}):
\ba M =
4\,\pi\,\left(\frac{r_v}{c_0}\right)^3\,\delta_0\,\rho_0\,\left[\ln(1+y)
-\frac{y}{1+y}\right],\label{M_NFW}\\
V^2 = 4\,\pi\, G\,\delta_0\,\rho_0\,\left(\frac{r_v}{c_0}\right)^2\,
\left[\frac{\ln(1+y)}{y}
-\frac{1}{1+y}\right],\label{Vsq_NFW}\ea
The total energy of the halo
(evaluated at the virial radius) is given by \cite{Mo}:
\begin{equation}
E_{_{\mathrm{NFW}}} = -\, G \, \frac{{M_v}^2}{2 r_v}\, F_0
\label{eq:E_nfw}
\end{equation}
where $F_0$ has the approximate values:
\begin{equation}
F_0=\frac{2}{3}+\left(\frac{c_0}{21.5}\right)^{0.7}\end{equation}

\section{SP and NFW comparison}

In order to compare both halo models, 
we have to make various physically motivated assumptions.
First, we want both models to describe a halo of the same size,
but since the virial radius, $r_v$, is the natural ``cut--off''
length scale at which the halo can be treated as an isolated
object in equilibrium, ``same size'' must mean same virial mass,
$M_v$, by equation (\ref{rvir}). 

Second. Both models must have the same maximum value for
the rotational velocity obtained from equations (\ref{rotvel}) and
(\ref{Vsq_NFW}). This is a plausible assumption, as it is based
on the Tully--Fisher relation, \cite{TF}, a very well established
result that has been tested successfully for galactic systems,
showing a strong correlation between the total luminosity of a
galaxy and its maximum rotational velocity. It can be shown that the
Tully--Fisher relation has a
cosmological origin \cite{Navarro,Jesus}\footnote{An article is
currently being prepared based on the results of \cite{Jesus}.}, associated with the 
primordial power spectrum
of fluctuations (the so called ``cosmological  Tully-Fisher
relation''), hence it is possible to translate the correlation
between maximum rotational velocity and total luminosity to a
correlation between maximum rotational velocity and total ({\it
i.e.} virial) mass. Since, by construction we are assuming that the
polytropic and NFW halos have the same $M_v$, their maximum
rotational velocities must also coincide. 

Our third assumption is
that the polytropic and NFW halos, complying with the
previous requirements, also have the same total energy evaluated at the
cut--off scale $r=r_v$, which can easily be computed for each model. The main
justification for this assumption follows from the fact that
the total energy is a fixed quantity in the collapse and subsequent
virialized equilibrium of dark matter halos~\cite{Padma1}.

Since all structural variables of the NFW halo depend only on the
virial mass, once we provide a specific value for $M_v$ all
variables become determined in terms of physical units by means
of equations (\ref{eq:c0}, \ref{rvir} and \ref{y}). Polytropic halos, on the other
hand, lack of a closed analytic expression for mass, velocity and
density profiles. In this case, equations (\ref{eq_ZMx}) (or
(\ref{eq:LE})) yield numerical solutions for these profiles
expressed in terms of the three free parameters
$\left\{\rho_c,\,\sigma_c,\,n\right\}$. The comparison of these
profiles with those of the NFW halos requires that we find
explicit values of these free parameters, so that the
conditions that we have outlined are met. Since we have selected
three comparison criteria for three parameters, we have a
mathematically consistent problem.

Following the guidelines described above, we proceed to compare
NFW and polytropic halos for $M_v$ ranging from $10^{10}M_{\odot}$ up to
$10^{15}M_{\odot}$. From the present comparison we find that the values for central
density, $\rho_c$, of the polytropic halos are inversely
proportional to $M_v$, while the values for the central
velocity dispersion, $\sigma_c$, are directly proportional to it
(this is expected, since $\sigma_c$ is a scale parameter in
self--gravitating systems). The polytropic index, $n$, is almost
constant for the selected range of $M_v$, showing a very small
growth as $M_v$ increases. This implies the same qualitative
behavior of the Tsallis parameter $q$: it is also almost constant
and is slowly increasing as the virial mass grows. It is
worthwhile mentioning that  the proportionally term $K_n$ in the
polytropic equation of state (\ref{eq:edo}) shows a very
noticeable change, rapidly growing as $M_v$ increases. All
these results are displayed explicitly in Table \ref{tab:t1}. 
The comparison between both models in mass profiles, velocity profiles and density profiles
is shown in figures \ref{fig:1}, \ref{fig:2} and left panel of figure \ref{fig:3} respectively.
SP and NFW models have both the same virial mass, $M_v = 10^{12}\,M_\odot$. 
For other values of $M_v$ the
mass, velocity and density profiles are qualitatively similar to
the ones displayed in these figures.

The right panel of figure \ref{fig:3} shows the
slope of the logarithmic density profile as a function of radius. On
the figure we can see more easily how the slope changes form the inner
to the outer regions of the halo. A generalized NFW profile like the
one presented at the beginning of section 3
(\cite{Austin,Dehnen,Hansen}) will lie between the SP and NFW in this figure.

\section{Including the galactic disk}

So far we have analyzed only the global structure of the dark matter halo
without considering the effects of the luminous galaxy within. If one wishes
to test a given model with observational results, it is necessary to add the
baryonic disk as a dynamical component of the model. We present 
some of the general ideas of a well known method used to establish a relation
between the dynamical properties of disk galaxies formed within virialized halos, 
with the properties of the dark matter halo itself \cite{Mo}. 

The initial conditions for the formation of the galactic disk are the following: an spherical
halo model (SP, NFW, etc.) with a given mass, velocity and
density profile; a total angular momentum $J$, total energy $E$ and 
a fraction $f_d$ of baryionic particles that will constitute the disk. In a dynamical 
context, the formation of disks inside dark matter halos can be well described 
with a model in which the halo responds adiabatically to the slow formation of the 
disk (due to energy dissipation of the baryonic particles). The method 
assumes that disk formation takes place in such a way that the final angular momentum 
of the disk, $J_d$, is a fraction of the total angular momentum of the halo:
$J_d=j_dJ$; this transfer of angular momentum is reasonable based
on a hypothesis of adiabatic ensemble of the disk. $J_d$ is
given by:
\begin{eqnarray}
J_d &=&2\pi\int_{0}^{r_v}rV_d(r)\Sigma_d(r)rdr, \nonumber \\
&=&M_dh_dV_v\int_{0}^{r_v/h_d}e^{-u}u^2 \frac{V_c(r)}{V_v}du,
 \label{momangdisk}
\end{eqnarray}
\noindent where $u=r/h_d$, $V_v$ is the virial velocity and $V_c(r)$ is
the total velocity profile(disk+halo).

The initial parameters of formation are given then by
the scale quantities: $M_v$, $r_v$ or $V_v$ (halo) and $f_d$
(disk), and the dynamical quantities: $\lambda=\frac{J|E|^{-1/2}}{GM_v^{5/2}}$ 
(halo) and $j_d$ (disk).

The central hypothesis of the method is that the angular momentum of each 
particle in the halo, before and after disk formation, is conserved (due to 
the adiabatic assumption). Thus, a particle that is initially at a mean 
radius, $r_i$, ends up at mean radius, $r$, both related by:
\begin{equation} 
M(r_i)r_i=M_f(r)r, \label{eq:cons}
\end{equation}

The halo evolves by several dynamical processes which end up with the baryonic 
particles collapsing into a plane to form a disk. The dark matter particles remain
in a spherically symmetric distribution, but they do feel a gravitational pull which
makes each mass shell of the dark halo to shrink from its original position.

The galactic system is thus formed by two components: an spherically
symmetric halo in virial equilibrium and a flattened disk with
azimuthal symmetry on centrifugal equilibrium, both characterized
by a radial density profile ($\rho(r)$ for the halo and
$\Sigma(r)$ for the disk). The observed surface brightness profiles for disk galaxies
favor an exponential form for the mass density of the baryonic disk:
$\Sigma(r)=\Sigma_{d,0}e^{-r/h_{d}}$, where $h_d$ is a scale radius, and 
$\Sigma_{d,0}$ is the central surface density. For this distribution, the mass 
profile is:
\begin{equation} \label{dmass}
M_{d}(r)=M_{d}\left[1-e^{r/h_{d}(1+r/h_{d})}\right],
\end{equation}
where $M_{d}=2\pi\Sigma_{d,0}h_{d}^2$.

The rotation curve for a disk infinitely flat is given by
Newtonian mechanics; a gravitational potential can be calculated
using Poisson's equation combined with the hypothesis of
centrifugal equilibrium and the mass distribution of the disk
(equation ($\ref{dmass}$)); the calculation yields:
\begin{equation} \label{vdisk}
V_{d}^{2}(r)=4\pi G\Sigma_{d,0}h_d
z^2\left[I_{0}(z)K_{0}(z)-I_{1}(z)K_{1}(z)\right],
\end{equation}
where $z=r/2h_d$. Since real galaxies are not completely flat, we
will correct the last formula according to an observational analysis
\cite{Burlak}, the correction diminish the value of
$V_d$ in $5\%$ approximately.

The final mass distribution of matter $M_f$ is the sum of the
dark matter inside the initial radius and the mass contributed by
the disk:
\begin{equation} \label{mf}
M_f(r)=M_d(r)+M(r_i)(1-f_d).
\end{equation}

The velocity profile of the halo is given by the virial
theorem:
\begin{equation} \label{VDM}
V_{DM}^2=G\left(\frac{M_f(r)-M_d(r)}{r}\right).
\end{equation}

The total rotational curve is the quadratic sum of disk and halo
contributions. The mass distribution and the circular velocity of the disk, 
$M_d(r)$ and $V_d(r)$, depend on $h_d$ and $\Sigma_{d,0}$, see 
(\ref{dmass},\ref{vdisk}). The scale radius is obtained in terms of the initial 
conditions and the formula for the angular momentum of the disk (combining the 
expression for the parameter $\lambda$, with (\ref{momangdisk})):
\begin{equation} \label{scaleradius}
h_d=\left(\frac{j_d}{f_d}\right)\lambda
G^{1/2}M_vr_v^{1/2}|E|^{-1/2}f_r
\end{equation}
where:
\begin{equation} \label{fr}
f_r\equiv\left[\int_0^{r_v/h_d}e^{-u}u^2\frac{V_c(h_du)}{V_v}\right]^{-1}
\end{equation}
The central surface density $\Sigma_{0,d}$ is obtained from the
Freeman relation:
\begin{equation} \label{sigma}
\Sigma_{0,d}=\frac{M_d}{2\pi h_d^2}=\frac{f_dM_v}{2\pi h_d^2}
\end{equation}

To calculate the final mass distribution $M_f$ we need to find
the initial radius of contraction, $r_i$, for each radius $r$,
this is done by combining (\ref{eq:cons}), (\ref{mf}) and
(\ref{dmass}) and solving the resulting transcendental equation.
Once we obtain $M_f$, the velocity profile of the dark matter
component can be calculated from equation (\ref{VDM}).

It is important to mention that in the end both dynamical components
(disk and halo) depend
on the value of the scale radius, $h_d$, which turns out to depend implicitly on
the final total circular velocity of the system (\ref{scaleradius},\ref{fr}), 
but $V_c$ also depends on $h_d$, so we need an iterative process to solve all the 
equations until a convergence for $h_d$ is obtained. Using the description presented,
the model is complete and can be used to obtain total rotational
curves from a set of initial conditions.

This method can be used for different mass profiles of
the halo, and allows us to obtain results for different halo
models; in particular we will use it for the NFW and SP models.

Both halo models which have been described before are
characterized by isotropic velocity distributions for dark matter
particles. However it is currently known that dark matter structures
are actually anisotropic (see for example \cite{Dehnen} and
\cite{Moore}, see also \cite{TDR}, and \cite{Lokas}). Such models are parametrized by
the quantity $\beta$, which is a measure of anisotropy:
$\beta=1-\sigma_{\theta}/\sigma_{r}$, where $\sigma_{\theta}$ and
$\sigma_{r}$ are the tangential and radial velocity dispersions. In the
simplest case of isotropic orbits, $\sigma_{\theta}=\sigma_{r}$ and
$\beta=0$. Analysis using numerical simulations in the past were not
very clear about the value of $\beta$: in \cite{Cole}, $\beta\sim0$
near the center and $\beta\sim0.2$ at the virial radius whereas in
\cite{Huss} $\beta\sim0.4$ at $r_v$. More recent works also suggest a
universal relation between $\beta$ and the logarithmic slope of the
radial density distribution \cite{Dehnen,Moore}.

In order to include the effects of anisotropy in our analysis we would
have to generalize the expressions for the different dynamical
profiles, both for the NFW and SP models, using a given model for
the radial dependence of $\beta$. Also, the model described for the
adiabatic contraction of the halo due to formation of the disk in the
halo would have to be generalized from circular to elliptical
orbits. Such analysis is out of the scope of the present
paper. 

However, it is important to say that realistic models for dark
matter halos should include the effects of anisotropy, then
the fact that in the Tsallis' formalism described here, the entropy is maximized for
isotropic systems tells us that such a description can not be the final
description of the nature of dark matter structures, our intention is
more modest, to propose it as alternative approximation to real
systems.

\section{Comparison with observations}

Now that we have described a dynamical model for the galactic system
(disk+halo), we can use it to compare the NFW and SP models against 
observations coming from a sample of real disk galaxies. 

The most direct approach would be to have observed rotation curves and 
surface brightness profiles (density distributions) for a representative sample of 
galaxies, then we could use a decomposition method to infer the components 
of the profile corresponding to dark and luminous matter for each galaxy in the
sample; in this way, we would be able to analyze the radial
distribution of each component and then, test both SP and NFW models to see
which one fits better with the observations. However, as noticed by \cite{JZ}, 
such a sample which is also complete in galactic properties is hard to 
find. We are interested in a sample of galaxies that could be representative
in the local Universe, i.e, that covers a wide range of luminosities, surface
brightness and morphological types. Having such a sample will allow us to
make an statistical analysis to test the SP and the NFW models. An alternative
method to make the comparison we want is based in using global parameters of 
galaxies and the relations between them to explore the mass amounts of 
luminous and dark matter in disk galaxies. This method favors the statistical
significance over precision. The approach used here will be on this direction,
and we will use the outline described by \cite{JZ} to make our analysis. 

In \cite{JZ}, the authors compiled a sample with the desired characteristics:
high quality surface photometry in the optical and
near-infrared band (to obtain density profiles and integral
luminosities), information about the rotation curves comes from
the $H_I$ or $H_{\alpha}$ line-widths, and the total $H_I$ gas
flux to take into account the gas contribution to the mass
distribution. This sample is
composed mainly by three subsamples described in 
\cite{deJong,deJong2,Verheijen,Bell}.

Some restrictions on the
original samples were made in order to obtain the final one
presented in \cite{JZ}. One of them is about the relative
inclination of the galaxies: the final sample of galaxies should
not be highly nor slightly inclined. Otherwise either the surface
brightness profiles in the former case, or the rotation velocities
in the latter one, are not reliable. At the end, only galaxies
with inclinations in the range of 35 to 80 degrees were considered.
Galaxies with clear signs of
interaction (mainly from the Verjeihen sample \cite{Verheijen}) and with rotation
curves which are still increasing at the last measured outer
radius were excluded from the sample. Only some LSB galaxies present this feature. Fortunately,
for most of the LSB galaxies in the sample the
synthetic rotation curves are reported, so it was possible to perform
such analysis in the exclusion criteria. The
final sample consists of 78 galaxies and is not complete
in a volume-limited sense, but is more or less homogeneous in the
range of basic parameters: luminosity, surface brightness and
morphological type. Dwarf galaxies were not included.

In order to have observations from the different samples as
uniform as possible, the raw data for each one of them was taken
an then corrected for the different factors that alter the actual
measure of a given parameter. The total magnitudes were corrected
for galactic extinction \cite{Schlegel},
$K-$correction \cite{Poggianti}, and internal extinction \cite{Tully2}. 
The surface brightness were corrected for galactic
extinction, $K-$correction, cosmological surface brightness
dimming, and inclination (geometrical and extinction effects). For
the latter correction the authors in \cite{JZ} followed the method
presented in \cite{Verheijen2} and
considered LSB galaxies as optically thin in all bands; they
defined LSB galaxies as those whose disk central surface
brightness in the $K-$band after correction is larger than 18.5
mag/arcsec$^2$. The 21 cm line-widths were corrected for
broadening due to turbulent motions and for inclination, following
\cite{Verheijen}.

The description of the sample, homogeneous data corrections and
transformations from observational to physical parameters is
carried on by the authors in \cite{JZ} and the reader is refereed to their
paper for more information about it.

With the information about the luminous disk structure parameters
that we have from the sample (which can be used to obtain $h_d$
and $\Sigma_{0,d}$), the velocity component $V_d(r)$ can be calculated
usin equation (\ref{vdisk}). A global
quantity associated with the disk will be the maximum of $V_d(r)$:
\begin{equation} \label{maxdisk}
V^2_{d,m} =4\pi GK\Sigma_{0,d}h_d
\end{equation}
where $K\approx0.193$; the maximum is located at
$r\approx2.2h_d$. The disk mass inside this radius is, according
to equation (\ref{dmass}): $M_d(2.2h_d)=0.64M_d$. Thus, without
introducing further assumptions, we may define the ratio of
maximum disk velocity to maximum total (or dynamical) velocity,
$V_{d,m}/V_{c,m}$, which is a global quantity that can be directly
compared with theoretical predictions.

The $V_{d,m}/V_{c,m}$ ratio is not defined at a unique radius, but
it can be related to the dynamical, total, disk mass ratio
$M_{dyn}/M_d$, defined at an specific radius. In particular at
radius $r_m$ where the total rotational curve reaches its maximum
we have:
\begin{eqnarray}
\left(\frac{M_{dyn}}{M_d}\right)_{r_m}&=&\frac{V_{c,m}^2r_m}{GM_d(r_m)} \nonumber \\
&\approx&\frac{V_{c,m}^2xh_d}{Gf_L2\pi\Sigma_{0,d}h_d^2}\propto\frac{x}{f_L}
\left(\frac{V_{d,m}}{V_{c,m}}\right)^{-2},
\end{eqnarray}
where we have used the virial theorem, the Freeman relation and
equation ($\ref{maxdisk}$). The important assumption made is that
$r_m$ is proportional to the scale radius: $r_m=xh_d$. In fact $x$
actually changes from galaxy to galaxy, depending on the
disk surface density and halo mass distribution ($x\approx2.2h_d$
for disk dominated galaxies, and $x>2.2$ for halo dominated
galaxies); $f_L$ is the fraction of the total disk mass at
$r_m$ and depends on the value of $x$ (when $x=2.2h_d$,
$f_L=0.64$); however it was shown by \cite{JZ} that $x$ changes slightly 
from galaxy to galaxy in the used sample (see figure 3a and 4a of \cite{JZ}).
The use of $V_{d,m}/V_{c,m}$ instead of $M_{dyn}/M_d$ is
suitable because it  can be obtained directly
from observational parameters, without the assumptions needed to
calculate $M_{dyn}/M_d$.  

One of the principal results obtained in \cite{JZ}
is that the ratio $V_{d,m}/V_{c,m}$ correlates mainly with
the disk surface density $\Sigma_d$. Therefore we will use this result to 
compare the NFW and the SP models with the observational results coming from
the sample\footnote{We note that there is still controversy 
regarding which is the main global property of galaxies that
correlates better 
with the ratio of dark to baryonic matter, some works support the results of 
\cite{JZ}, see for example \cite{Kar}, and others claim that the 
luminosity is the key parameter, see for example \cite{Sal1,Sal2,Sal3}.}.

To do the comparison, we need to give the values for the free parameters of
the models. $M_v$, $r_v$ and $V_v$ will be the same in both
models, to account for same ``size'' of the halos, and they must
have the same disk fraction $f_d$, meaning that the galactic disk
inside the spherical halos will be of the same mass; we also
require the same value for $j_d$, in this way the fraction of
angular momentum of the disk to the halo will not depend, on a
dynamical level, on the mass profile that the galactic system has
before disk formation.

We will take $M_v=1\times10^{12}M_{\odot}$, which is a characteristic value 
for the mass of dark matter halos, as the standard value
for the virial mass. For simplicity, we will assume that $f_d$ and
$j_d$ have the same values for all disks (that is, independent of
$\lambda$); it turns out that in fact, the value of $f_d$ can not
vary significantly from galaxy to galaxy, because this will
predict larger scatter in the Tully-Fisher than the actually observed 
(Undergraduate Thesis \cite{Jesus}; an article which contains these an
other results is currently in preparation). 
We will also use the standard assumption in modeling
disk formation: $j_d/f_d=1$ \cite{Fall,Mo}; although it is unclear if this 
hypothesis is
appropriate (it depends on the efficiency of disk formation) we
will take it as valid for the purpose of this work, arguing that
the change of this assumption does not change the statistical trend
of the presented results. It has not been established an
appropriate value for the disk fraction $f_d$. However, a
plausible upper limit is the baryon fraction of the Universe as a
whole ($f_B=\Omega_B/\Omega$), taking a Big Bang nucleosynthesis
value for the baryonic density, $\Omega_B=0.015h^{-2}$ \cite{Walker} 
gives an upper limit value on $f_d$ of 0.1
for a $\Lambda CDM$ cosmological model (see also \cite{Mo}); however, 
the efficiency in forming the disk could be quite
low, implying that $f_d$ could be substantially lower than $f_B$,
this is confirmed by the results in \cite{Jesus} where an analysis 
on the scatter of the TF
relation showed that $f_d$ has a numerical value even less than
$0.05$. Taking into account these considerations we will take as
average values for all galaxies in both models: $f_d=j_d=0.03$.

In figure \ref{fig:4} we show a plot of the ratio $V_{d,m}/V_{c,m}$ against $\Sigma_d$
for the described sample of galaxies (open circles). Despite the large
scatter, an almost
linear trend can be seen between this quantities, HSB (high
surface brightness) galaxies (corresponding approximately to
values of log$\Sigma_d$ greater than 2.5) have greater values of
$V_{d,m}/V_{c,m}$ than LSB (low surface brightness) galaxies. This
means that the luminous matter content is greater for HSB than for
LSB galaxies. The shown picture is consistent with a well known
result: LSB galaxies are dark matter dominated systems within
optical radius.

The value of the graphic is that it allows us to bound
statistically the possible values of the $V_{d,m}/V_{c,m}$ ratio
that galaxies with a given surface density can have. As was proved by 
\cite{JZ}, the size of this range of values (associated with dispersion
on the graphic) is due to different disk mass, {\it i.e.} different virial
mass, color and morphological type that galaxies with the same
$\Sigma_d$ have. As has been said, the value of $f_d$ can not
vary significantly from galaxy to galaxy, nevertheless increasing
$f_d$ will increase the value of $V_{d,m}/V_{c,m}$ and viceversa, as
an example, taking an interval of 0.02 around the central value of $f_d=0.03$ changes the value of
the ratio $V_{d,m}/V_{c,m}$ in 4\% on average, being less for HSB
galaxies and larger for LSBs. In figure 2 of \cite{JZ} the result of
varying $f_d$ can be seen explicitly.
The observed dispersion seen in figure \ref{fig:4} mainly dependent on total 
disk mass  and integral color (see figure 7a of \cite{JZ}) has a cosmological
origin which is explained with virial mass and concentration dispersion on the 
halos respectively (a galaxy formed inside a halo with a large value of the 
concentration parameter, will have an integral color redder than one formed 
inside a halo with a smaller value, an analogous conclusion is found for halos 
with different masses).

Figure \ref{fig:4} also shows clearly that NFW models can not reproduce at
satisfaction the results obtained for the compiled sample without
introducing unrealistic values for the virial mass or the
concentration parameter. This is one of the results that lead us
to the possibility of seeking an alternative to the NFW paradigm.
The curves shown in figure \ref{fig:4}
represent both models with average values for their respective
parameter; it's clear from the figure that the SP model follows in better 
agreement the average
behavior of the observational sample than the NFW model. In terms
of the parameters $\rho_c$, $\sigma_c$ and $n$ of the stellar
polytrope model, the scatter of the observational results is
explained in the same way as for the NFW model, $\rho_c$ takes the
role of the concentration parameter and $\sigma_c$ the one
corresponding to the virial mass (we found that the polytropic index 
$n$ is almost constant for virial masses
in the range $10^{10}M_{\odot}$ to $10^{15}M_{\odot}$).

In \cite{JZ}, the same sample of observational data was also
compared with results coming from self-consistent galaxy evolution
models (see figure 3 of \cite{JZ}). Such semi-numerical models were
developed in \cite{Avila,Firmani}; they follow the disk galaxy
formation and evolution in a hierarchical $\Lambda$CDM
scenario. These models include self-consistently halo formation and
evolution, disk star formation and feedback processes, the
gravitational dragging of the halo due to disk formation (where the
adiabatic invariance is generalized to elliptical orbits), secular
bulge formation and other evolution processes. The overall main
difference is a small increase on the $V_{d,m}/V_{c,m}$ ratio in
figure (\ref{fig:4}) for the more realistic
models compared to the simpler model described in this paper (see
section 3.2 of \cite{JZ}), the difference is more notorious for HSB
galaxies. It is remarkable then that despite the lack of real features
like anisotropy, star formation and feedback, the simple (disk+halo) models
presented above offer reasonable results compared to the
evolutionary models, at least in what respect to figure \ref{fig:4},
which is enough for the purpose of our work.

\section{Conclusions}

Motivated by the fact that stellar polytropes are the equilibrium states in
Tsallis' non--extensive entropy formalism, we have found the structural
parameters of those stellar polytropes that allows us to compare them with NFW
halos of virial masses in the range $10^{10}<\Mvir/M_\odot<10^{15}$. The
criteria for this comparison consists in demanding that the polytrope describes
a halo having the same virial mass, virial radius, maximum rotational velocity 
and total energy as the NFW halo. These three conditions are sufficient to
determine the three structural parameters
$\left\{\rho_c,\,\sigma_c,\,n\right\}$ of the polytropic model; 
the results are displayed in Table \ref{tab:t1}.

\begin{figure*}
\centering
\includegraphics[height=7cm]{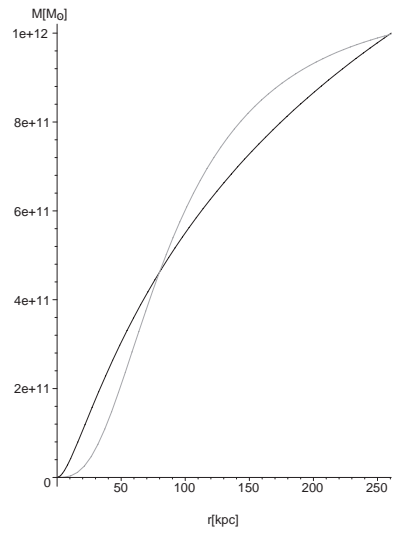}
\caption{Mass  profiles for a NFW halo with
$M_v=10^{12}\,M_\odot$ and $r_v=260$ kpc (solid curve) and
compared fit stellar polytrope (dashed curve).}
\label{fig:1}
\end{figure*}

\begin{figure*}
\centering
\includegraphics[height=7cm]{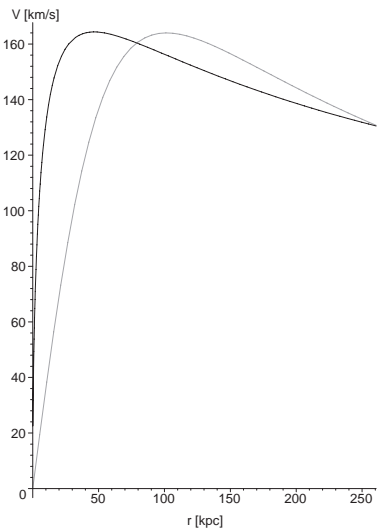}
\caption{Velocity profiles for the same NFW halo of figure 1 (solid line) and
its compared stellar polytrope (dashed curve).}
\label{fig:2}
\end{figure*}

\begin{figure*}
\centering
\includegraphics[height=7cm,width=13cm]{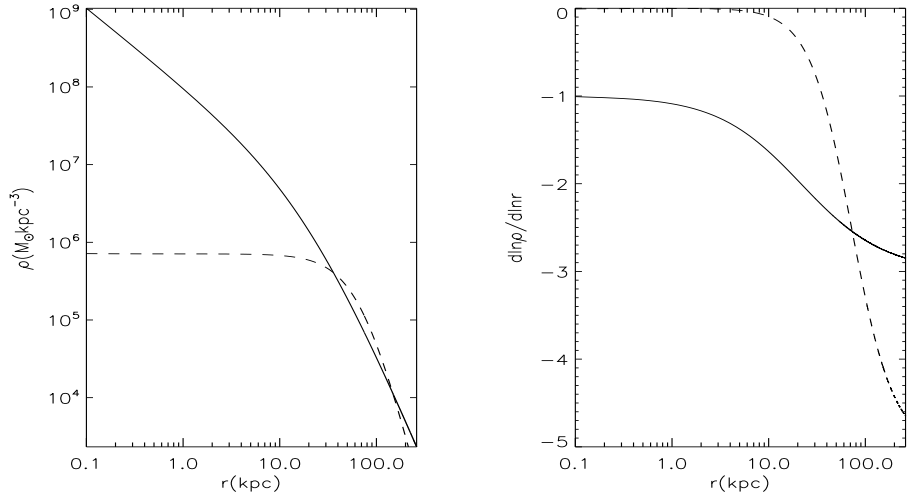}
\caption{In the left panel the density profiles for the same NFW halo of figures 1 and
2 (solid curve) and its compared stellar polytrope (dashed curve). In
the right panel, the value for the slope of the logarithmic density
profile is explicitly shown for different radii.}
\label{fig:3}   
\end{figure*}

We emphasize that the criteria which determine these
polytropes are based on physically motivated
assumptions: the virial radius and mass are the natural
parameters characterizing the size of a given halo, same maximum
velocity follows from the Tully--Fisher relation, while same total
energy follows from the virilization process. As shown in
Figure \ref{fig:1}, the mass distribution of the polytrope grows
much slower than that of the NFW halo up to a large radius (100
kpc) containing the core and the region where visible matter
concentrates. Hence, as shown by Figure \ref{fig:2}, the velocity
profile of the polytrope is much less steep in the same region
than that of the NFW halo. These features are consistent with the
fact that NFW profiles predict more dark matter mass
concentration than what is actually observed in a large sample of
galaxies~\cite{vera,Bin,JZ}. Also, as shown
in the left panel of figure \ref{fig:3}, the obtained polytropes have flat cores, very
similar to the flat isothermal cores observed in LSB galaxies (as
a contrast, the cuspy cores of NFW halos seem to be at odds with
these observations~\cite{cdm_problems_1, cdm_problems_2,
cdm_problems_3}, see also \cite{vera, Bin}). This flat density
core is a nice property, which combined with reasonable mass and
velocity profiles, qualifies these polytropes as reasonable
(albeit idealized) models of halo structures.

\begin{figure*}\centering
\includegraphics[height=7cm]{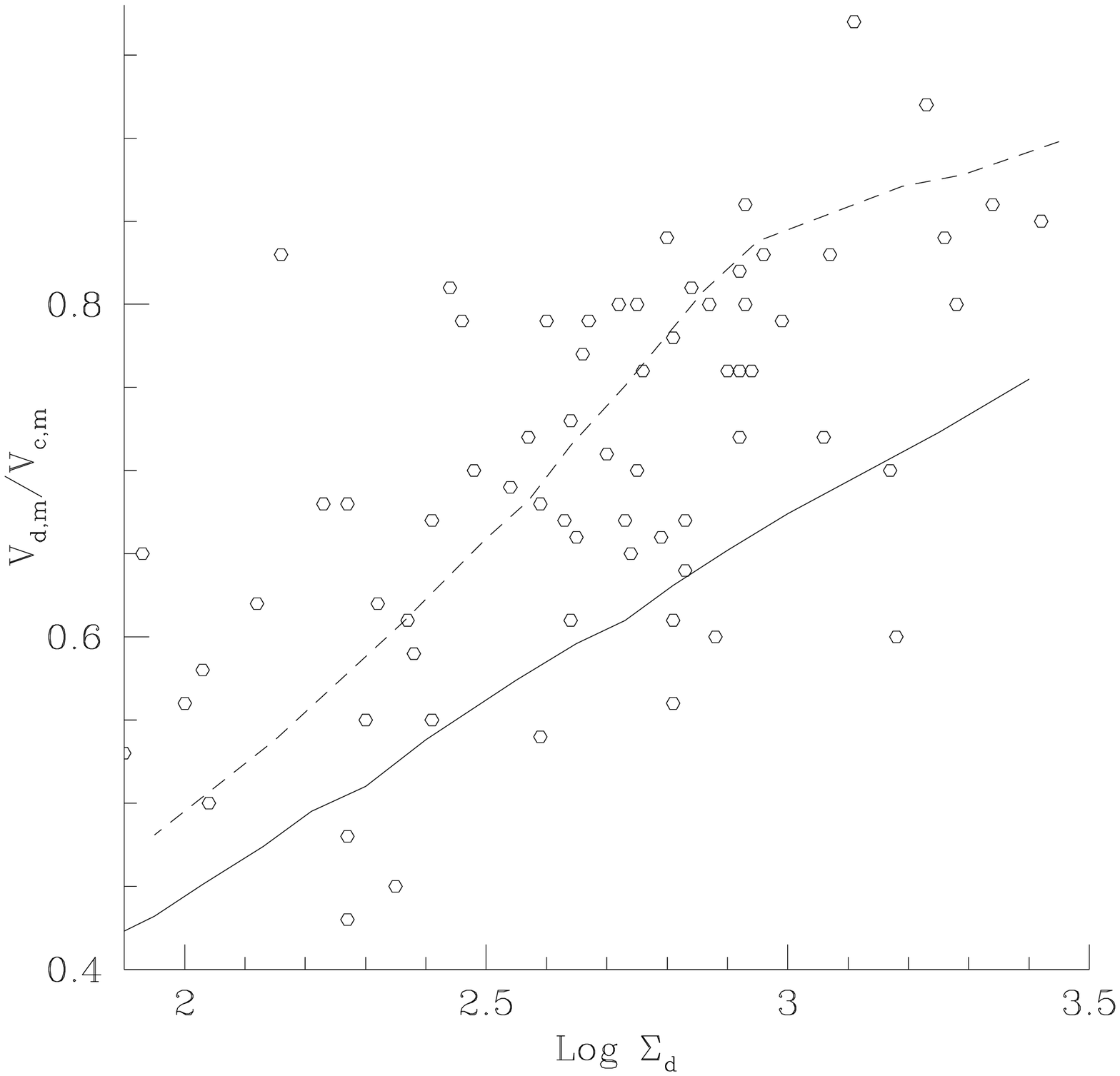}
\caption{Luminous to total dark matter content vs central surface density; 
open circles correspond to observational data, the solid line represents the
NFW model and the dashed one the SP model. Both models have $M_v=10^{12}M_{\odot}$.}
\label{fig:4}
\end{figure*}

However, in spite of their nice theoretical properties ({\it
i.e.} their connection to Tsallis' formalism) and reasonable
similarity with equivalent NFW halos, the stellar polytropes we have
examined are very idealized configurations. Thus, we are not
claiming that they provide a realistic description of halo
structures. Instead, we suggest that their described features and their
connection with Tsallis' formalism might
indicate that the latter could yield useful information to
understand the evolution and virialization processes of dark
matter. Although it is necessary to pursue this idea by
means of more sophisticated methods, including the use of
numerical simulations along the lines pioneered by \cite{TS2},
the simple approach we have presented has already given
interesting results. For example, with respect to the parameter $q$( 
we recall that it is a free parameter of the Tsallis' non-extensive 
thermodynamics and which has not been fixed for the cosmological case), in 
this work we were able to
determine its behavior as a function of the virial mass, and turns
out to be almost constant, with values around $q\approx 1.2$.
This result could be used in other contexts where extended
statistical mechanics is also applied \cite{Tsallis2}. We have also shown that the SP model
is favored over the NFW model regarding the dark matter content in disk 
galaxies (within the optical radius) which is shown by the average behavior of 
the observational sample in figure \ref{fig:4}. These results show that 
the NFW halo model can be enhanced with the
use of alternative paradigms in statistical mechanics, which seems to  
solve a recurrent item which throws a shadow in such an excellent description
as the NFW model is.

As mentioned, the results presented in this work show that
a dark matter halo made out of particles which satisfy a polytropic 
equation of state, describes the halo in a way that is comparable with the 
description obtained from the NFW numerical simulations. 
Furthermore, our description has the advantage of not having a cuspy
density profile near the galactic center. These results do not 
directly imply that dark  matter halos obey a non-extensive entropy
formalism. Further tests and experiments are 
needed in order to consider that such formalism is the one describing the 
dynamics of actual dark matter halos. At the moment, this
idea is only a suggestion which is reinforced by our analysis. 

Finally, we would like to address the possibility of a future analysis of the
actual statistical mechanics treatment ruling the structure
formation process of dark matter halos by using numerical simulations
to obtain the equation of state obeyed by such systems at different
times of their evolution and
then determine towards which equilibrium state are they moving to. Thus 
allowing us to have a more reliable method to know whether or not the Tsallis' formalism
is appropriate for describing such self-gravitating systems.

\ack 
We are grateful to Vladimir Avila-Reese for his comments and suggestions 
to the manuscript of the present work. We also like to thank the
anonymous referee for valuable suggestions and comments. We acknowledge partial support by CONACyT
M\'exico, under grants SEP-2004-C01-47209-F, 32138-E and 34407-E, and DGAPA-UNAM IN117803,
IN113206-2 grants.
JZ acknowledges support from DGEP-UNAM and CONACyT scholarships.


\pagebreak


%
\begin{table}
\caption{Parameters characterizing the polytropes while being
compared to NFW halos}
\begin{tabular}{cccccccc}
\br
$\log_{10}(M_v/M_{\odot})$ & $\rho_c\, [M_{\odot}/\rm{pc}^3]$ &
$\sigma_c\, [\rm{Km/s}]$  & $n$ & $q$ & $K_n$ & $v_{\rm{max}}\,
[\rm{Km/s}]$ & $r_v\, [\rm{kpc}]$
\\ \hline 15 & $3.7 \times 10^{-4}$ & $982$ & $4.93$ & $1.29$ &
$4873.4$ & $1504$ & $2606.2$\\
12 & $7.5 \times 10^{-4}$ & $108$ & $4.87$ & $1.30$ &
$478.94$ & $164$ & $260.6$\\
11 & $9.0 \times 10^{-4}$ & $52$ & $4.83$ & $1.30$ &
$221.82$ & $79.1$ & $120.9$\\
10 & $1.2 \times 10^{-3}$ & $25$ & $4.82$ & $1.30$ &
$100.68$ & $38.2$ & $56.1$\\
\br
\end{tabular}
\label{tab:t1}
\end{table}

%

\section*{References}

\begin{thebibliography}{30}

\bibitem{Tsallis}Tsallis C, Nonextensive statistics: Theoretical, experimental and 
computational evidences and connections, 1999 {\it Braz J Phys} {\bf 29} 1
\bibitem{PL}Plastino A R and Plastino A, Stellar polytropes and Tsallis' entropy, 
1993 {\it Physics Letters A} {\bf 174} 384
\bibitem{TS1}Taruya A and Sakagami M, Gravothermal Catastrophe and Generalized 
Entropy of Self-Gravitating Systems, 2002 {\it Physica A} {\bf 307} 185 
[cond-mat/0107494]
\bibitem{TS2}Taruya A and Sakagami M, Long-term Evolution of Stellar Self-Gravitating 
System away from the Thermal Equilibrium: connection with non-extensive statistics, 
2003 {\it Phys. Rev. Lett.} {\bf 90} 181101; See also Taruya A and Sakagami M, 
Self-gravitating Stellar Systems and Non-extensive Thermostatistics, 
2004 {\it Continuum Mechanics and Thermodynamics} {\bf 16} 279-292 [cond-mat/0310082]
\bibitem{NFW}Navarro J F, Frenk C S and White S D M, A Universal Density 
Profile from Hierarchical Clustering, 1997 {\it ApJ} {\bf 490} 493 [astro-ph/9611107]
\bibitem{Blok1}de Blok W J G, McGaugh S S, Rubin V C, High-resolution rotation curves 
of LSB galaxies: Mass Models, 2001 {\it ApJ} {\bf 122} 2396 [astro-ph/0107366] 
\bibitem{Blok2}de Blok W J G, McGaugh S S, Bosma A and Rubin V C, Mass Density 
Profiles of LSB Galaxies, 2001 {\it ApJ} {\bf 552} L23 [astro-ph/0103102]
\bibitem{Bin}Binney J J and Evans N W, Cuspy
Dark-Matter Haloes and the Galaxy, 2001 {\it MNRAS} {\bf 327} L27 [astro-ph/0108505]
\bibitem{B-S}Borriello A and Salucci P, The dark matter distribution in disc galaxies,
2001 {\it MNRAS} {\bf 323} 285 [astro-ph/0001082]
\bibitem{B-O}Blais-Ouellette, Carignan C and Amram P,
Multiwavelength Rotation Curves to Test Dark Halo Central Shapes, 2002 
{\it Preprint} astro-ph/0203146
\bibitem{Bosma3}Bosma A, Invited review at IAU Symposium 220: {\it Dark
Matter in Galaxies}, Sydney Australia, 21-25 Jul 2003, {\it Preprint} astro-ph/0312154
\bibitem{self}Spergel D N and Steinhardt P J, Observational evidence for 
self-interacting cold dark matter, 2000 {\it Phys. Rev. Lett.} {\bf 84} 3760
[astro-ph/9909386]
\bibitem{warm}Colin P, Avila-Reese V, and Valenzuela O,
Substructure and halo density profiles in a Warm Dark Matter scenario, 2001
{\it ASP Conference Series} {\bf 230} p. 651-652 [astro-ph/0009317]  
\bibitem{DMe1}Guzm\'{a}n F S and Matos T, Scalar fields as dark matter in spiral galaxies, 
2000 {\it Class. Quantum Grav.} {\bf 17} L9 [gr-qc/9810028]
\bibitem{DMe2}Matos T, Guzm\'{a}n F S and N\'{u}\~{n}ez D,
Spherical Scalar Field Halo in Galaxies, 2000 {\it Phys. Rev. D} {\bf 62} 061301 
[astro-ph/0003398]
\bibitem{DMe3}Matos T and Guzm\'{a}n F S, On the Space Time of a Galaxy,
2001 {\it Class. Quantum Grav.} {\bf 18} 5055 [gr-qc/0108027]
\bibitem{DMe4}Matos T and Ure\~{n}a-L\'{o}pez L A,
Scalar Field Dark Matter, Cross Section and Planck-Scale Physics, 2002 
{\it Phys. Lett. B} {\bf 538} 246 [astro-ph/0010226]
\bibitem{DMe5}Guzm\'{a}n F S and Ure\~{n}a-L\'{o}pez L A, Evolution of the 
Schr\"odinger--Newton system for a self--gravitating scalar field, 2004 
{\it Phys. Rev. D} {\bf 69} 124033 
\bibitem{Ruffini}Ruffini R and Bonazzola S, Systems of selfgravitating 
particles in general relativity and the concept of equation of state, 
1969 {\it Phys. Rev. D} {\bf 187} 1767

\bibitem{Pointe}Pointecouteau E, Arnaud M and Pratt G W, The
structural and scaling properties of nearby galaxy clusters. I. The universal mass profile, 
2005 {\it A\&A} {\bf 435} 1 [astro-ph/0501635]
\bibitem{Voigt}Voigt L M and Fabian A C, Galaxy cluster mass profiles, 
2006 {\it Submitted to MNRAS} [astro-ph/0602373]
\bibitem{Zapa}Zappacosta L, Buote D A, Gastaldello F, Humphrey P
J, Bullock J, Brighenti F and Mathews W,  The absence of adiabatic
contraction of the radial dark matter profile in the galaxy cluster a2589, 
2006 {\it Submitted to ApJ} [astro-ph/0602613]

\bibitem{Tsallis1}Cabral-Rosetti L G, Matos T, Nunez D, Sussman R and
Zavala J, Empirical testing of Tsallis entropy in density profiles of
galactic dark matter halos, 2006, in preparation.
\bibitem{B-T}Binney J and Tremaine S, {\it Galactic dynamics}, 1987,
ed. Princeton University


\bibitem{TsaPla}Tsallis C, Mendes R S and Plastino A R, The role
of constraints withing generalized nonextensive statistics, 
1998 {\it Physica A} {\bf 261} 534
\bibitem{Taru_new}Taruya A and Sakagami M, Gravothermal
catastrophe and Tsallis' generalized entropy of selfgravitating
systems. 3. Equilibrium structure using normalized q values, 
2003 {\it Physica} {\bf 322} 285 [cond-mat/0211305]
\bibitem{Taru_05}Taruya A and Sakagami M, Antonov problem and
quasi-equilibrium states in N-body system, 
2005 {\it MNRAS} {\bf 364} 990 [astro-ph/0509745]
\bibitem{Lima}Lima J A S and de Souza R E, Power law stellar distributions, 
2005 {\it Physica A} {\bf 350} 303 [astro-ph/0406404]
\bibitem{Hansen_Tsallis}Hansen S H, Egli D, Hollenstein L and
Salzmann C, Dark matter distribution function from non-extensive statistical mechanics, 
2005 {\it New Astronomy} {\bf 10} 379 [astro-ph/0407111]


\bibitem{NT}Taylor J E and Navarro J F, The Phase-Space
Density Profiles of Cold Dark Matter Halos, 
2001 {\it ApJ} {\bf 563} 483 [astro-ph/0104002]
\bibitem{Sto}Stoehr, F, Circular velocity profiles of dark matter haloes, 
2006 {\it MNRAS} {\bf 365} 147
\bibitem{Austin}Austin C G, Williams L L R, Barnes E I, Babul A
and Dalcanton J J, Semianalytical Dark Matter Halos and the Jeans Equation, 
2005 {\it ApJ} {\bf 634} 756 [astro-ph/050657]
\bibitem{Dehnen}Dehnen W and McLaughlin D, Dynamical insight into dark-matter haloes, 
2005 {\it MNRAS} {\bf 363} 1057 [astro-ph/0506528]
\bibitem{Hansen}Hansen s H and Stadel J, The velocity anisotropy - density slope relation, 
2005 [astro-ph/0510656]

\bibitem{Mo}Mo H J, Mao S and White S D M, The Formation of Galactic Disks,
1998 {\it MNRAS} {\bf 295} 319 [astro-ph/9707093]
\bibitem{LoHo}Lokas E L and Hoffman Y, Nonlinear evolution of spherical perturbation 
in a non-flat Universe with cosmological constant, 2001 {\it Preprint} astro-ph/0108283. 
See also Lokas E L, Structure formation in the quintessential Universe, 2001 
{\it Acta Phys.Polon.} {\bf B32} 3643
\bibitem{JZ}Zavala J, Avila-Reese V, Hern\'andez-Toledo H and Firmani C, 
The luminous and dark matter content of disk galaxies, 2003 {\it A\&A} {\bf 412} 633
[astro-ph/0305516]
\bibitem{c0}Eke V R, Navarro J F and Steinmetz M, The Power Spectrum Dependence of 
Dark Matter Halo Concentrations, 2001 {\it ApJ} {\bf 554} 114 [astro-ph/0012337]
\bibitem{TF}Tully R B and Fisher J R, A new method of determining distances to galaxies, 
1977 {\it A\&A} {\bf 54} 661, see also Colin P, Avila-Reese V and Valenzuela O, 
Substructure and halo density profiles in a Warm Dark Matter Cosmology, 2000 
{\it ApJ} {\bf 542} 622 [astro-ph/0004115]

\bibitem{Navarro}Steinmetz, M and Navarro J F, The Cosmological
Origin of the Tully-Fisher Relation, 1999 {\it ApJ} {\bf 513} 555 [astro-ph/9808076]

\bibitem{Jesus}Zavala J, {\it El plano fundamental luminoso y
bari\'onico de las galaxias de disco}, 2003, Undergraduate thesis, UNAM.
\bibitem{Padma1}Padmanabhan T, {\it Structure formation in the universe}, 1993,
Cambridge University Press.
\bibitem{Burlak}Burlak A N, Gubina V A and Tyurina N V, Allowance for the effect of finite 
thickness of galactic disks on the circular rotation velocity, 1997 {\it Astro. Lett} 
{\bf 23} 522

\bibitem{Moore}Hansen S H, Moore B, A universal density slope -
velocity anisotropy relation for relaxed structures, 
2006 {\it New Astronomy} {\bf 11} 333 [astro-ph/0411473]
\bibitem{TDR}Matos T, Nunez D and Sussman R A, The Spacetime
associated with galactic dark matter halos, {\it Gen. Rel. Grav.},
{\bf 37} 769 [astro-ph/0402157]
\bibitem{Lokas}Lokas E L and Mamon G A, Properties of spherical
galaxies and clusters with an nfw density profile, 
2001 {\it MNRAS} {\bf 321} 155 [astro-ph/0002395]
\bibitem{Cole}Cole S and Lacey C G, The Structure of dark matter
halos in hierarchical clustering models, 
1996 {\it MNRAS} {\bf 281} 716 [astro-ph/9510147]
\bibitem{Huss}Huss A, Jain B and Steinmetz M, The Formation and
evolution of clusters of galaxies in different cosmologies, 
1999 {\it MNRAS} {\bf 308} 1011 [astro-ph/9703014]
 
\bibitem{deJong}de Jong R S and Van der Kruit P C, Near-infrared and optical broadband 
surface photometry of 86 face-on disk dominated galaxies. I. Selection, observations and data 
reduction, 1994 {\it A\&ASS} {\bf 106} 451 
\bibitem{deJong2}de Jong R S, Near-infrared and optical broadband surface photometry of 
86 face-on disk dominated galaxies. III. The statistics of the disk and bulge parameters, 
1996 {\it A\&A} {\bf 313} 45 [astro-ph/9601005]
\bibitem{Verheijen}Verheijen M W A and Sancisi R, The Ursa Major Cluster of 
Galaxies. IV : HI synthesis observations, 2001 {\it A\&A} {\bf 370} 765 [astro-ph/0101404]
\bibitem{Bell}Bell E, Barnaby D, Bower R G, de Jong R S, Harper Jr.
D A, Hereld M, Loewenstein R F and Rauscher BJ, The star formation
histories of low surface brightness galaxies, 2000 {\it MNRAS} {\bf 312} 470 [astro-ph/9909401]
\bibitem{Schlegel}Schlegel D J, Finkbeiner D P and Davis M, Maps of Dust Infrared Emission 
for Use in Estimation of Reddening and Cosmic Microwave Background Radiation Foregrounds,
1998 {\it ApJ} {\bf500} 525 [astro-ph/9710327]
\bibitem{Poggianti}Poggianti B M, K and evolutionary corrections from UV to IR, 
1997 {\it A\&ASS} {\bf 122} 399 [astro-ph/9608029]
\bibitem{Tully2}Tully R B, Pierce M J, Huang J, Saunders W, Verheijen M A W and Witchalls P L, 
Global Extinction in Spiral Galaxies, 1998 {\it AJ} {\bf 115} 2264 [astro-ph/9802247]
\bibitem{Verheijen2}Verheijen M A W, PhD Theses, Groningen University,
1997 
\bibitem{Kar}Karachentsev I D, Scatter of SC Galaxies in the Tully-Fisher Diagram and 
the Dark Matter Problem, 1991 {\it SvAL} {\bf 17} 367
\bibitem{Sal1}Salucci P and Borriello A, {\it The Distribution of Dark Matter in Galaxies: 
Constant-Density Dark Halos Envelope the Stellar Disks}, 
2001 in "Proceedings of the International Conference DARK 2000", Heidelberg, Germany, p.12
\bibitem{Sal2}Salucci P and Persic M, Maximal Halos in High-Luminosity Spiral Galaxies,
1999 {\it A\&A} {\bf 351} 442 [astro-ph/9903432]
\bibitem{Sal3}Salucci P, Ashman K M and Persic M, The dark matter content 
of spiral galaxies, 1991 {\it ApJ} {\bf 379} 89
\bibitem{Fall}Fall S M and Efstathiou G, Formation and rotation of disc galaxies 
with haloes, 1980 {\it MNRAS} {\bf 193} 189
\bibitem{Walker}Walker T P, Steigman G, Kang H S, Schramm D M and Olive K A, 
Primordial nucleosynthesis redux, 1991 {\it ApJ} {\bf 376} 51 

\bibitem{Avila}Avila-Reese, V, Firmani C an Hern\'andez X, On the
Formation and Evolution of Disk Galaxies: Cosmological Initial
Conditions and the Gravitational Collapse, 1998 {\it ApJ} {\bf 505} 37
\bibitem{Firmani}Avila-Reese, V and Firmani C, Properties of Disk
Galaxies in a Hierarchical Formation Scenario, 
200 {\it RevMexAA} {\bf 36} 23 [astro-ph/0001403]

\bibitem{vera}McGaugh S S, Rubin V C and de Blok W J G, High-Resolution Rotation 
Curves of Low Surface Brightness Galaxies. I. Data, 2001 {\it ApJ} {\bf 122} 2381
[astro-ph/0107326]
\bibitem{cdm_problems_1}Moore B, The Nature Of Dark Matter, 1994 {\it Nature} {\bf 370} 629
[astro-ph/9402009]
\bibitem{cdm_problems_2}Flores R and Primack J P, Observational and Theoretical 
Constraints on Singular Dark Matter Halos, 1994 {\it ApJ} {\bf 427} L1 [astro-ph/9402004]
\bibitem{cdm_problems_3}Burkert A, {\it The structure of dark matter halos. 
Observation versus theory}, Aspects of Dark  Matter in Astro-and Particle Physics (ed.
Klapdor-Kleingrothaus H V and Ramachers Y), 1997 {\it Preprint} astro-ph/9703057
\bibitem{Tsallis2}Tsallis C, {\it Nonextensive Statistical Mechanics and its
Applications}, 2001, (Eds. Abe S and Okamoto, Y. Springer, Berlin, 2001)
%
\end{thebibliography}
\end{document}